\documentclass[prl,aps,twocolumn,showpacs,superscriptaddress]{revtex4}
\usepackage{graphicx}

\begin{document}
\title{The tail of the contact force distribution in static
granular materials}

\author{Adrianne R.T. van Eerd}
\affiliation{Condensed Matter and Interfaces, Utrecht University,
P.O. Box 80.000, 3508 TA Utrecht, The Netherlands}

\author{Wouter G. Ellenbroek} \affiliation{Instituut--Lorentz,
Universiteit Leiden, Postbus 9506, 2300 RA Leiden, The
Netherlands}

\author{Martin van Hecke}
\affiliation{Kamerlingh Onnes Lab, Leiden University, PO box 9504,
2300 RA Leiden, The Netherlands}

\author{Jacco H. Snoeijer}
\affiliation{School of Mathematics, University of Bristol,
University Walk, Bristol BS8 1TW, United Kingdom}

\author{Thijs J.H. Vlugt}
\affiliation{Condensed Matter and Interfaces, Utrecht University,
P.O. Box 80.000, 3508 TA Utrecht, The Netherlands}

\date{\today}

\begin{abstract}
We numerically study the distribution $P(f)$ of contact forces in
 frictionless bead packs, by averaging over the ensemble of
all possible force network configurations. We resort to umbrella
sampling to resolve the asymptotic decay of  $P(f)$ for large $f$,
and determine $P(f)$ down to values of order $10^{-45}$ for
ordered and disordered systems in two and three dimensions. Our
findings unambiguously show that, in the ensemble approach, the
force distributions decay much faster than exponentially: $P(f)
\sim \exp(-f^{\alpha})$, with $\alpha\!\approx\!2.0$ for 2D systems,
and $\alpha\!\approx\!1.7$ for 3D systems.
\end{abstract}


\maketitle

The contact forces inside a static packing of grains are organized
into highly heterogeneous force networks, and can be characterized
by the probability density of contact forces $P(f)$~\cite{grm}.
Such force statistics were first studied in a series of
experiments that measured forces through imprints on carbon paper
at the boundaries of a granular assembly. Unexpectedly, the
obtained $P(f)$ displayed an exponential rather than a Gaussian
decay for large forces~\cite{carbon}. After these
initial findings, other experimental techniques have revealed
similarly exponentially decaying distributions of the boundary
forces ~\cite{noren,corwin}.

The first model that captured this exponential decay was the
pioneering $q$-model, where scalar forces are balanced on a
regular grid~\cite{coppersmith}. Later studies found, however,
that the nature of the tail of $P(f)$ depends on the details of
the stochastic rules for the force transmission in this model and
need not be exponential~\cite{bouchaud}. 
Other explanations for the exponential tail hinge on ``entropy maximization''~\cite{kruit}, or closely related, on an analogy with the Boltzmann distribution~\cite{rottler,metzger}. The essence of the latter argument is that a uniform sampling of forces that (i) are all positive (corresponding to the repulsive nature of contact forces), and (ii) add up to a constant value (set by the requirement that the overall pressure is constant), strongly resembles the microcanonical ensemble, in which configurations are flatly sampled under the constraint of fixed total energy.

As it is difficult to experimentally access contact forces {\em
inside} the packing, many direct numerical simulations of $P(f)$
have been undertaken~\cite{pf1,pf2}. While numerous of these
studies claim to find an exponential tail as well, the evidence is
less convincing than for the carbon paper experiments: apart
from~\cite{pf1}, nearly all numerical force probabilities bend
down on a logarithmic plot, suggesting a faster than exponential
decay~\cite{pf2}. In addition, new experimental techniques using
photoelastic particles~\cite{bob} or emulsions
\cite{brujic,dinsmore}, have produced bulk measurements, and these
also reveal a much faster than exponential decay for $P(f)$,
consistent with a Gaussian tail.

These contradicting findings completely reopen the discussion on
the tail of $P(f)$. The presently available data for $P(f)$ have
been obtained from a wide variety of systems and models, and
parameters such as dimensionality, hardness of grains, 
bulk vs boundary measurements, may ultimately 
all play a role in determining the asymptotics of $P(f)$. 
In addition, in many cases, the true asymptotic nature of $P(f)$ 
is hard to probe, since obtaining reliable data for forces much larger 
than the average value $\langle f \rangle$ remains a challenge.

In this paper we will probe the tail of $P(f)$ in the force
network ensemble \cite{prlensemble,PRET,worm,unger,tighe,panja}. 
We numerically resolve the probability for large forces using the 
technique of {\em umbrella sampling}~\cite{umbrella}, which yields 
accurate statistics for $P(f)$ for relative probabilities down to $10^{-45}$ 
and $f$ up to $f=15 \langle f \rangle$. 
This high accuracy is crucial for excluding any cross-over effects~\cite{metzger}, 
and allows to unambiguously identify the behaviour for $f \gg \langle f \rangle$. 
We study frictionless
systems in two and three dimensions, both with ordered and
disordered contact networks, and also explore the effect of system
size and contact number. 

For all these systems, we have found that the ensemble yields a much faster 
than exponentially decaying force distribution. 
The dimensionality of the system is crucial, while other factors hardly affect the asymptotics: 
$P(f)$ decays as $\exp(-f^\alpha)$, with $\alpha\!=\!2.0\pm0.1$ in two dimensions, while in three
dimensions $\alpha\!=\!1.7\pm0.1$~\cite{foothertz}.

\paragraph{Force network ensemble and umbrella sampling ---} 
The ensemble approach to force networks is inspired by the
proposal of Edwards to assign an equal probability to all
``blocked" states, {\em i.e.} states that are at mechanical
equilibrium~\cite{edwards}. By limiting the Edwards ensemble to a single contact
geometry of a packing of frictionless spheres~\cite{houches}, where the contact
forces are the remaining degrees of freedom and all allowed
force-configurations are sampled with equal weight, we obtain the
force network ensemble. We restrict ourselves to spherical particles with
frictionless, repulsive contacts, so that every contact force
$f_i$ corresponds to one scalar degree of freedom. Furthermore, we
require all $f_i \geq 0$ due to the repulsive nature of the
contacts. As the equations of mechanical equilibrium are linear in
the contact forces, one can cast the solutions
$\vec{f}=(f_1,f_2,\cdots)$ in the form $\vec{f} = \vec{f}_0 +
\sum_k c_k \vec{v}_k$. The solution space is spanned by the
vectors $\vec{v}_k$ and $\vec{f}_0$, and can be sampled through
the coefficients $c_k$ -- for details we refer to
Refs.~\cite{prlensemble,PRET,tighe}. For a hexagonal packing
(two dimensional), these vectors are easily constructed using so-called
``wheel moves''~\cite{tighe}, but for other packings we have
obtained $\vec{v}_k$ and $\vec{f}_0$ from a simulated annealing
procedure~\cite{prlensemble}. Ensemble averages using a uniform
measure in this force space then become

\begin{equation}
\langle q \rangle = \Omega^{-1} \int_{\mathcal{C}} d\vec{c} \, q~,
\quad \quad \Omega \equiv \int_{\mathcal{C}} d\vec{c}~,
\end{equation}
where the integral runs over the coefficients $c_k$ limited to the
convex subspace $\mathcal{C}$ for which all $f_i \ge
0$~\cite{unger}.

To obtain accurate statistics for large forces we perform umbrella sampling.
The idea is to bias the numerical
sampling towards solutions with large forces, using a 
Monte Carlo technique with a modified measure
$d\vec{c}\,\rho(\vec{c})/\Omega$, and then correct for this bias
when performing the averages, $\langle q \rangle=\langle q /\rho
\rangle_{\rm umbrella}$, since

\begin{equation}
\langle q \rangle = 
\Omega^{-1} \int_{\mathcal{C}} d\vec{c} \,
\rho(\vec{c}) \, \left( \frac{q}{\rho(\vec{c})} \right)~.
\end{equation}
Defining $f_{\rm max}$ as the largest force for a given $\vec{c}$,
we have used a measure $\rho(\vec{c})\propto e^{W(f_{\rm max})}$,
where $W$ is chosen such that the probability of $f_{\rm max}$ in
the modified ensemble is approximately flat in the range $\langle
f\rangle < f_{\rm max} < 15\langle f\rangle$. 
We have verified that this procedure exactly reproduces $P(f)$ in
the range accessible by the conventional unbiased
sampling~\cite{tighe}. 
However, forces of the order of $15\langle f \rangle$ are now sampled 
only $10^{4}$ times less frequently than around $\langle f \rangle$, even 
though their relative probability is about $10^{-45}$, leading to the 
spectacular improvement in the numerical accuracy~\cite{futurepaper}.

\begin{figure}[tbp]
\includegraphics[width=8.5cm]{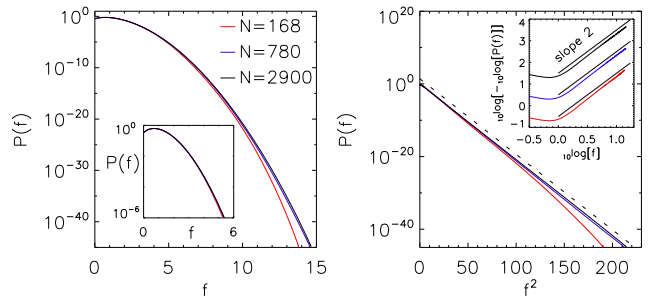}
\centering \caption{Force probabilities in two dimensional,
hexagonal packings with periodic boundary conditions and
increasing system size as indicated by the number of particles
$N$. (a) $P(f)$ decays much faster than exponentially, and rapidly
converges to its asymptotic form with $N$. The inset illustrates
that system size effects are hardly visible for $P(f)$ down to
$10^{-6}$. (b) $\log P$ vs $f^2$ becomes a perfectly straight line
for large systems, indicating that the tail of $P(f)$ is well
described by a Gaussian decay $\sim \exp(-f^2)$ (dashed line). The
inset shows that on a triple-log plot, the asymptotic decay
attains a slope close to 2, confirming the Gaussian tail (see
text). Curves are offset for clarity, and lines are guides to the
eye.} \label{fig1}
\end{figure}

{\em Hexagonal packings in two dimensions ---} A well studied
geometry for which the force network ensemble yields nontrivial results is
when all particles are of equal size and form a hexagonal lattice
\cite{prlensemble,PRET,tighe}. The umbrella sampling allows us,
for the first time, to access the statistics beyond $f=5\langle f
\rangle$. Figure~\ref{fig1}(a) shows that $P(f)$ decays much faster
than exponentially, and that effects of the finite size of the
system are weak. Figure~\ref{fig1}(b) illustrates that for
increasingly large systems, $P(f)$ rapidly converges to an
asymptotic form which is characterized by a purely Gaussian decay.
This can also be seen in the inset of Fig.~1(b), where we exploit
the fact that we have access to $P(f)$ over more than forty
decades: Assuming that for large $f$, $P(f)\sim
\exp(-(f/\lambda)^\alpha)$, one can infer the exponent $\alpha$ 
from the asymptotic slope of a triple-log plot in which $\log(-\log P)$ 
is plotted as function of $\log f$ (base 10)~\cite{corwin}. 
In Fig.~\ref{fig1}(c) we find $\alpha$ to be $2.0 \pm 0.1$, 
confirming that the tail of $P(f)$ is well described by a 
Gaussian decay~\cite{footsmallf}.

\begin{figure*}[t]
\includegraphics[width=17.8cm]{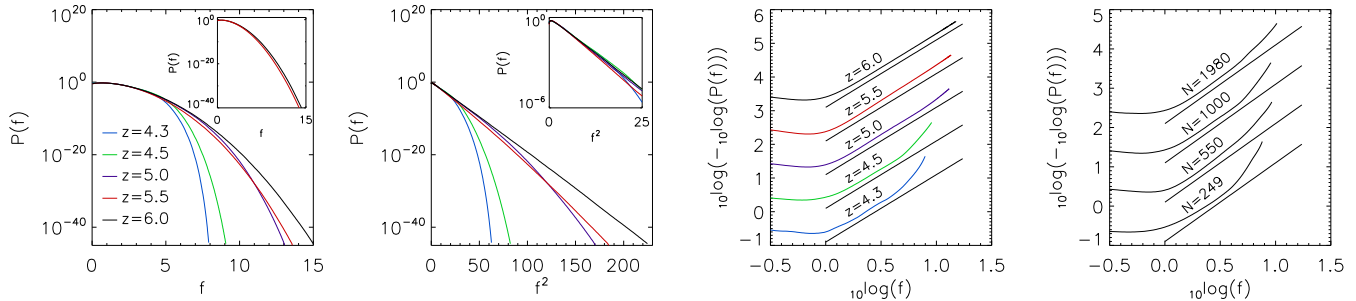}
\centering \caption{Force distribution for two dimensional
systems. (a) Effect of contact number $z$ on $P(f)$ for disordered
packings of $N=1000$ particles. The inset compares the $P(f)$ for a
disordered packing with $z=6$ and $N=1000$ and the hexagonal
packing for $N=2900$. (b) The same data as in (a), now plotted as
$\log P(f)$ vs $f^2$, tends to a straight curve for large $z$. The
inset shows that on a smaller range, all curves look Gaussian. (c)
Same data as in (a-b), now on a triple log plot. The range in $f$
over which $P(f)$ looks Gaussian grows with contact number $z$.
(d) For fixed small $z=4.5$, $P(f)$ appears to approach a Gaussian
tail for large $N$. } \label{fig2}
\end{figure*}

{\em Disordered packings in two dimensions --- } To investigate
the effect of packing disorder and coordination number $z$, we
have created packings from molecular dynamics simulations of soft
particles in periodic boundary conditions (see
\cite{prlensemble,worm}). The coordination number $z$ is
controlled by the pressure in the simulations. Once a packing is
obtained, its geometry is kept fixed, and we subsequently explore
the ensemble of force networks for these packings.

\begin{figure*}[btp]
\includegraphics[width=17.8cm]{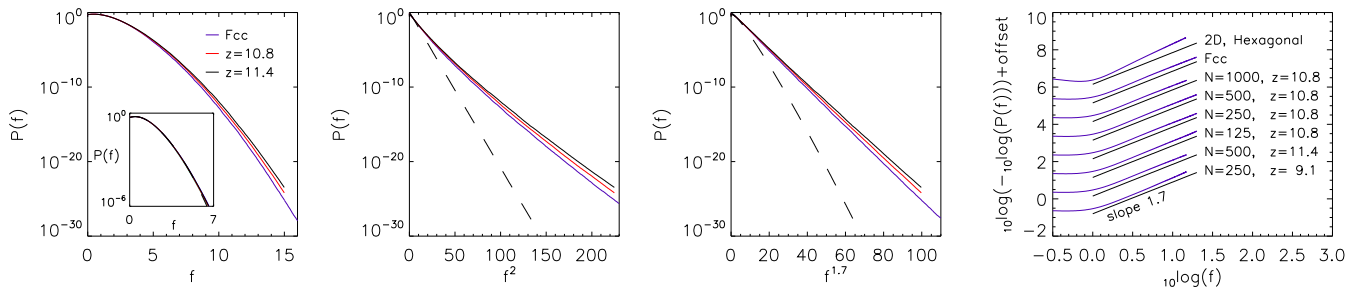}
\centering \caption{Force distribution for three dimensional
systems. (a) $P(f)$ for two disordered and a regular fcc packing
of $N=500$ particles. (b) Same, now plotted as function of $f^2$.
The dashed line corresponds to a hexagonal packing in 2D, which
has a Gaussian tail - the tail of $P(f)$ for 3D systems is
significantly less steep. (c) Same data, now plotted as function
of $f^{1.7}$ - the tails for the $P(f)$ of 3D packings are now
straight. (d) The change from 2 to 1.7 is also clearly visible in
the triple log plot. For a range of system sizes and contact
numbers, we robustly find that $P(f)\sim \exp(-(f/\lambda)^\alpha)$
with an exponent $\alpha \approx 1.7$ for 3D systems --- for
comparison we also show the Gaussian distribution for the 2D
hexagonal packing. Note that for small systems and small contact
number ($N=250,z=9.1$), finite size deviations, similar to those
observed in two dimensions, can be seen.
 } \label{fig3}
\end{figure*}

For all 2D disordered packings, $P(f)$ decays much faster than
exponentially, as shown in Fig.~\ref{fig2}. Comparing the ordered
hexagonal packings to a disordered system with equal coordination
number, $z=6$, we find nearly indistinguishable $P(f)$ (inset
Fig.~2(a)). This suggests that the packing (dis)order is not
important for $P(f)$.

The contact number influences the asymptotics of $P(f)$: a lower
$z$ leads to a faster decay (Fig.~\ref{fig2}(a-b)), although in
the restricted range $f<5 \langle f \rangle$,
the force distribution appears very close to Gaussian for all $z$
(inset Fig.~\ref{fig2}(b)). For the lowest $z$ in particular, this
tendency is cut off at large $f$, which can be clearly seen in the
triple-log plot Fig.~\ref{fig2}(c), where all curves tend towards
a well-defined slope $\alpha=2$ for intermediate $f$ but cross
over to a much faster decay for large $f$. We suggest that this is
a finite size effect, which is most severe when $z$ approaches the
isostatic point ($z=4$), where there are less and less degrees of
freedom available~\cite{isostatic,worm}. Indeed, data for $z=4.5$ and
increasing system sizes suggest that the ``kink'' in the triple
log plots becomes less severe for large systems (Fig.~\ref{fig2}(d))
--- our data are not conclusive as to whether this kink will
disappear for $N\rightarrow \infty$.

In conclusion, for two dimensional, frictionless systems, the
ensemble approach yields $P(f)$ that have a tail with decays at
least as fast as a Gaussian.

{\em Three dimensional packings --- } We now turn to three
dimensional systems, which again have been generated using 
molecular dynamics. Similar to what happens in two dimensions,
Fig.~\ref{fig3}a shows that $P(f)$ decays faster than
exponentially, and disordered and regular (fcc) packings have very
similar force distributions. However, the decay is now {\em slower}
than Gaussian and much more accurately described by $P(f)\sim
\exp(-(f/\lambda)^\alpha)$ with an exponent $\alpha = 1.7 \pm 0.1$, 
see Fig.~\ref{fig3}(b-d). This exponent has been determined from the
triple-log plots of Fig.~\ref{fig3}(d) for a range of contact
numbers and system sizes, and in all cases the slope is close to
$\alpha= 1.7$ over a decade.

For comparison we have, in Fig.~\ref{fig3}(b-d), also included the
result for the hexagonal pack, which is seen to decrease
significantly more rapidly than the $P(f)$'s of the three
dimensional systems. Surprisingly, we thus find that the
dimensionality of the packing determines the nature of the tail of
$P(f)$.

\begin{figure}[tbp]
\includegraphics[width=8.5cm]{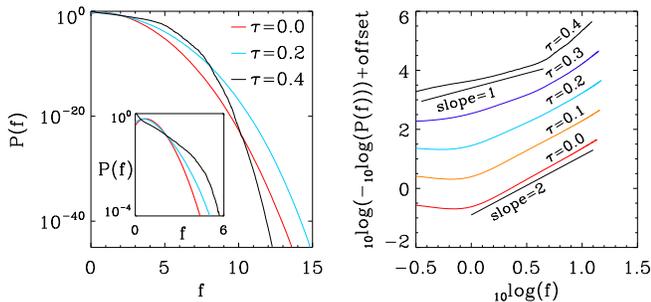}
\centering \caption{Two dimensional disordered system with $z=5.5$
experiencing a shear stress $\tau \equiv \sigma_{xy}/\sigma_{xx}$
\cite{worm}. (a) While for large $\tau$, the tail of $P(f)$ viewed
over a limited range broadens and may appear exponential (inset),
the asymptotic decay of $P(f)$ for $f>10$ in fact becomes steeper
(main panel). (b) The same point is illustrated in the triple log
plots, which also show data for $\tau=0.1$ and $0.3$. }
\label{fig.shear}
\end{figure}

{\em The effect of shear stress ---} From experiments on (two dimensional) sheared
packs of photoelastic grains, it was found that the distribution
broadened significantly, and developed an exponential-like regime
in a range up to $4\langle f \rangle$~\cite{bob}. The ensemble
indeed reproduces this qualitative feature for packs under shear.
As can be seen in Fig.~\ref{fig.shear}, however, there does not
seem to be a simple asymptotic decay. This is because the force
anisotropy induced by the shear stress yields a variation in
$\langle f \rangle$ depending on the orientation of the
contact~\cite{tighe,worm}. The total $P(f)$ becomes a
superposition over all orientations, of mixed force statistics,
and hence lacks a single characteristic feature.

\paragraph{Discussion ---}

In none of the cases we investigated, $P(f)$ exhibits an
exponential tail. 
The ``Boltzmann'' type arguments based on conservation of total 
force, which was suggested to be responsible for the exponential tail~\cite{rottler,metzger}, 
could in principle have been applicable here: 
indeed, all contact forces in the ensemble are 
positive and add up to a value proportional to the pressure. 
This reasoning, however, does not take into account that forces 
have to balance on each grain. 
Our results underline the importance of these additional constraints, 
which completely alter the properties of $P(f)$. 

The force distributions obtained in the ensemble are consistent 
with those obtained in most experimental and numerical studies. 
In contrast, experiments that measure forces at the boundaries 
appear to find an exponential decay of $P(f)$~\cite{carbon,noren,corwin}. 
This remains a crucial issue for the understanding of static granular media, 
since the force statistics provides insight to the proper measure to weigh 
the microscopic configurations corresponding to a macroscopic experimental protocol~\cite{corwin}. 
At present, one still lacks a relation between characteristics 
of the system (presence of boundaries~\cite{jacco_wallstuff}, 
the relative hardness of grains and boundaries~\cite{footnote}, 
friction and nonsphericity, ...) and the force distribution. 
Within the ensemble theory it would be possible address the effect of torque balance, 
which is clearly important for real (frictional) systems or even for
frictionless non spherical grains --- both friction and nonsphericity 
can in principle be included in the ensemble. Having seen that the normal
force balance conditions have different effects in two or three
dimensions, it would be very interesting to see whether torque
balance could yet again change the nature of the tail. 

W.G.E acknowledges financial support from the physics foundation FOM, 
M.v.H. and T.J.H.V. from NWO/VIDI, and 
J.H.S from a Marie Curie European Fellowship FP6 (MEIF-CT-2006-025104).


\begin{thebibliography}{99}
\bibitem{foothertz} Similar exponents emerge for the potential energy of 
Hertzian contacts~\cite{corwin}, which scale as $f^2$ and $f^{5/3}$ respectively. 
As the ensemble considers rigid particles without any contact law, 
this resemblance is a coincidence and can not explain our findings.
\bibitem{footsmallf} Intriguingly, the force distribution is surprisingly well approximated at small $f$ by $P(f)\simeq \frac13+\frac34f$, which hints that a simple analytic expression may exist for the hexagonal packing. A decent fit over the whole range of $f$ is given by $P(f)=(a+bf)e^{-c(f-f_0)^2}$, where $a,b$ are determined by the observed small $f$ behavior and $f_0,c$ are determined by $\langle f \rangle=1$ and $\int P(f)df=1$, but deviations from the numerical $P(f)$ can be observed in the tail.
\bibitem{footnote} Note that the glass beads used in the boundary measurements are
much harder than the particles used in the
bulk measurements of~\cite{bob,brujic,dinsmore}, and that realistic
particle rigidity is difficult to achieve in molecular dynamics.
\end{thebibliography}
\end{document}